\documentclass[sigconf]{acmart}
\AtBeginDocument{%
  }

\copyrightyear{2025} 
\acmYear{2025} 
\setcopyright{acmlicensed}
\acmConference[MM '25]{Proceedings of the 33rd
ACM International Conference on Multimedia}{October 27--31, 2025}{Dublin,
Ireland}
\acmBooktitle{Proceedings of the 33rd ACM International Conference on
Multimedia (MM '25), October 27--31, 2025, Dublin, Ireland}
\acmDOI{10.1145/3746027.3754921}
\acmISBN{979-8-4007-2035-2/2025/10}

\usepackage{enumitem}

\begin{document}

\title{Music2Palette: Emotion-aligned Color Palette Generation via Cross-Modal Representation Learning}
\author{Jiayun Hu}
\affiliation{%
  \department{School of Computer Science and Technology}
  \institution{East China Normal University}
  \city{Shanghai}
  \country{China}}
\email{51265901035@stu.edu.ecnu.cn}
\orcid{0009-0001-1606-5832}

\author{Yueyi He}
\affiliation{%
 \department{School of Software Engineering}
  \institution{East China Normal University}
  \city{Shanghai}
  \country{China}}
\email{10225101555@stu.ecnu.edu.cn}
\orcid{0009-0002-7550-0918}

\author{Tianyi Liang}
\affiliation{%
  \department{School of Computer Science and Technology}
  \institution{East China Normal University}
  \city{Shanghai}
  \country{China}}
\email{tyliang@stu.ecnu.edu.cn}
\orcid{0000-0001-8372-8379}

\author{Changbo Wang}
\affiliation{%
  \department{School of Computer Science and Technology}
  \institution{East China Normal University}
  \city{Shanghai}
  \country{China}}
\email{cbwang@cs.ecnu.edu.cn}
\orcid{0000-0001-8940-6418}

\author{Chenhui Li}
\authornote{Corresponding author.}
\affiliation{%
  \department{School of Computer Science and Technology}
  \institution{East China Normal University}
  \city{Shanghai}
  \country{China}}
\email{chli@cs.ecnu.edu.cn}
\orcid{0000-0001-9835-2650}

\renewcommand{\shortauthors}{Jiayun Hu, Yueyi He, Tianyi Liang, Changbo Wang, and Chenhui Li}

\begin{abstract}
Emotion alignment between music and palettes is crucial for effective multimedia content, yet misalignment creates confusion that weakens the intended message. However, existing methods often generate only a single dominant color, missing emotion variation. Others rely on indirect mappings through text or images, resulting in the loss of crucial emotion details. To address these challenges, we present Music2Palette, a novel method for emotion-aligned color palette generation via cross-modal representation learning. We first construct MuCED, a dataset of 2,634 expert-validated music-palette pairs aligned through Russell-based emotion vectors. To directly translate music into palettes,  we propose a cross-modal representation learning framework with a music encoder and color decoder. We further propose a multi-objective optimization approach that jointly enhances emotion alignment, color diversity, and palette coherence.  
Extensive experiments demonstrate that our method outperforms current methods in interpreting music emotion and generating attractive and diverse color palettes. Our approach enables applications like music-driven image recoloring, video generating, and data visualization, bridging the gap between auditory and visual emotion experiences. 
\end{abstract}

\begin{CCSXML}
<ccs2012>
   <concept>
       <concept_id>10010147.10010257.10010293.10010319</concept_id>
       <concept_desc>Computing methodologies~Learning latent representations</concept_desc>
       <concept_significance>300</concept_significance>
       </concept>
   <concept>
       <concept_id>10010405.10010469.10010474</concept_id>
       <concept_desc>Applied computing~Media arts</concept_desc>
       <concept_significance>300</concept_significance>
       </concept>
   <concept>
       <concept_id>10010147.10010257.10010293.10010294</concept_id>
       <concept_desc>Computing methodologies~Neural networks</concept_desc>
       <concept_significance>300</concept_significance>
       </concept>
 </ccs2012>
\end{CCSXML}

\ccsdesc[300]{Computing methodologies~Learning latent representations}
\ccsdesc[300]{Applied computing~Media arts}
\ccsdesc[300]{Computing methodologies~Neural networks}

% \vspace{-2em}
\keywords{Affective computing; color palette generation; music emotion recognition; multimodal representation}
% , transformer-based generative model
% \begin{figure}
%   \includegraphics[width=\textwidth]{figs/teaser.pdf}
%   % \Description{Enjoying the baseball game from the third-base
%   % seats. Ichiro Suzuki preparing to bat.}
% \end{figure}


\maketitle

\begin{figure}[!t]
	\centering
	\includegraphics[width=0.85\linewidth]{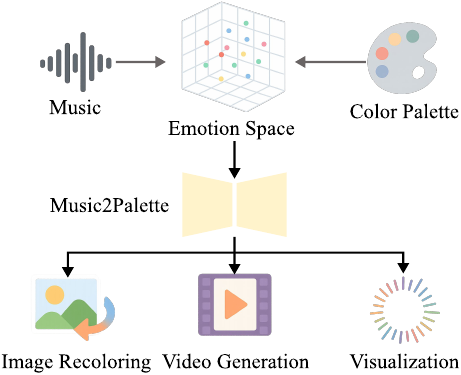}
	\caption{Music2Palette is an end-to-end model trained to generate palettes from music by aligning both modalities in a shared emotion space. By learning from matched music and palette pairs, we can create fitting color palettes for tasks like image recoloring, video generation, and visualization.
}
	\label{Pipeline}
\end{figure}

\section{Introduction}
Humans are inherently emotional beings. With the rapid development of digital media and social platforms,
people express their feelings through music, images, and videos~\cite{Zhao2020AnEV, Zhao2018PredictingPI}. Among these modalities, music and color, which influence the auditory and visual systems respectively, are especially effective in conveying emotion~\cite{TZaatar2023TheTP, Jonauskait2025DoWF, Palmer2013MusiccolorAA}. However, when the visual tone of an image or video does not match the background music, it can lead to emotion inconsistency and reduce the overall expressiveness of the content~\cite{Boltz2004TheCP, Hauck2022CrossmodalCB}. 
Building emotion-aligned palettes directly from music provides a promising way to guide the emotion tone of visual content and reduce emotion mismatch.
Such music-driven color palettes can serve as references in a variety of applications, such as image recoloring~\cite{Xu2025AestheticMI}, video generation~\cite{Liu2023EmotionAwareMD}, and user interface design~\cite{Shi2023DeStijlFG, Labrecque2012ExcitingRA}, enabling even novice creators to produce emotionally harmonious content.

Recently, emotion recognition has received increasing attention across modalities such as text~\cite{Nandwani2021ARO, Zhang2018DeepLF}, image~\cite{Yamamoto2021ImageER, Yang2022ExploitingEC}, and music~\cite{Li2024MusicER, Yang2012MachineRO}.
In cross-modal settings, many methods aim to align these modalities through shared emotion representations~\cite{Zhao2020EmotionBasedEM}.
However, few studies focus on music and color palettes. 
Some works attempt to map music to color, but they only predict one main color~\cite{hsiao2017methodology, Pesek2017TheMD}, which simplifies the dynamic emotion of music.
Other methods introduce intermediate modalities such as text or images. The text-based methods convert music emotions into descriptive labels~\cite{doh2023lp}, then use text-to-color tools like Text2Colors~\cite{DBLP:conf/eccv/BahngYCPWMC18} or Text2Palette~\cite{lei2021text2palette} to generate palettes.
The image-based methods retrieve emotionally aligned images~\cite{Zhao2020EmotionBasedEM} and extract dominant colors from those images~\cite{ahmed2020generalized, DBLP:journals/tog/Finkelstein15}.

However, these indirect methods often result in emotion loss, as the rich affective signals in music may not fully transfer across modalities. Moreover, it is difficult to align different modalities because they describe information in different ways and have features of different sizes. These issues often limit the diversity of the generated palettes and make the emotional expression less accurate. Currently, there are no suitable datasets or models designed to directly map music to palettes, and little work explores this direct and accurate emotional correspondence.

To address the lack of direct mapping between music and palettes and the loss of emotion information, we propose Music2Palette, a novel framework that produces emotion-aligned color palettes from music via cross-modal representation learning.
We construct a large-scale cross-modal dataset, MuCED, including 2634 high-quality music–palette pairs. By introducing emotion vectors, we map both music and palettes into Russell’s circumplex model, and use vector similarity to perform automatic emotion alignment. Expert evaluation is further employed to ensure strong emotion coherence and aesthetic quality in the curated pairs. To model the cross-modal affective mapping, we design a Transformer-based generative model that includes a music encoder and a color decoder, with added Gaussian noise to enhance diversity. 
To address the challenges of limited variation and emotion inconsistency in palette generation, we introduce a multi-objective loss function that optimizes color accuracy, palette diversity, and emotion consistency.
Extensive quantitative experiments indicate that our method outperforms existing baselines across affective coherence, aesthetic quality, and color diversity. Its strong performance in real-world scenarios, such as image recoloring, video generation, and data visualization, further highlights its broad utility and scalability across diverse modalities.

% In summary, the main contributions of our work are as follows:
In summary, our main contributions are as follows:
\begin{itemize}[topsep=0pt, itemsep=0pt]
\item We construct MuCED, the first extensive cross-modal dataset for music-to-palette generation. In this dataset, both music and palettes are described using emotion vectors based on Russell’s circumplex model. 
To ensure quality, we also include expert refinements to make sure the emotion is well-aligned, the palettes maintain high perceptual fidelity, and the results generalize across cultures.
\item 
We propose Music2Palette, an end-to-end cross-modal representation learning framework that encodes music into a shared latent emotion space and decodes it into semantically and perceptually compatible color palettes.
The joint optimization of color accuracy, diversity, and emotion alignment enables Music2Palette to generate palettes that are both visually expressive and affectively coherent.
\item We propose a set of evaluation metrics covering color diversity and emotion consistency, including Convex Hull Overlap and Bhattacharyya Coefficient for color distribution, as well as Emotion Similarity and Jensen-Shannon Divergence for emotion alignment. Experiments show that our method outperforms existing methods across these dimensions.
\end{itemize}

% \vspace{-4em}
\section{Related Work}
\subsection{Music Emotion Recognition and Representation}
Emotion recognition is commonly approached in two ways: categorical and dimensional.
Categorical emotion state (CES) models define a set of basic, discrete emotions, such as Ekman’s six basic emotions~\cite{Ekman1992AnAF}, Mikel’s eight emotional categories~\cite{Mikels2005EmotionalCD} and Geneva Emotion Music Scale~\cite{Zentner2008EmotionsEB}. 
Dimensional emotion space (DES) models represent emotions on continuous axes, most notably the Valence-Arousal-Dominance (VAD) model~\cite{Schlosberg1954ThreeDO}, where valence describes pleasantness, arousal indicates intensity, and dominance reflects a sense of control. As dominance is often unstable to measure, many studies focus only on valence and arousal~\cite{Hanjalic2006ExtractingMF, Kim2017BuildingEM, Zhao2017ContinuousPD}.
Based on this simplified two-dimensional space, Russell’s circumplex model~\cite{Russell1980ACM} organizes eight representative emotions in a circular layout according to their valence and arousal levels. In our work, we adopt these eight emotions to describe both music and color palettes. This shared representation captures the emotional meaning in both modalities and supports effective cross-modal alignment.

To represent music in an emotion-aware shared space, researchers have explored various music emotion recognition (MER) methods.
Early approaches rely on traditional machine learning models such as SVMs~\cite{hsiao2017methodology, Kim2010StateOT, Xianyu2016SVRBD}, using handcrafted features based on acoustic and music theory knowledge.
With the rise of deep learning, models gain the ability to learn directly from raw audio. CNNs~\cite{Han2016DeepCN, Mao2014LearningSF, huang2025dual} and LSTMs~\cite{Cai2016MaxoutNF, Chen2017MultimodalML, Grekow2021MusicER} have been widely used to capture both short-term sound patterns and how emotion changes over time.
More recently, Transformer-based models like the Audio Spectrogram Transformer (AST)~\cite{gong2021ast} have achieved strong results in music emotion tasks by modeling long-range dependencies. 
However, it is not designed for cross-modal alignment and cannot map emotion signals to visual modalities. To address this, we build a cross-modal framework based on AST, projecting music into a shared emotional space to enable direct alignment between music and color palettes.
% In our work, we adopt AST as the music encoder to project audio into a shared latent space, allowing alignment between the emotional signals in music and visual color palettes.

\subsection{Color Palette Emotion Recognition and Generation}
The connection between colors and emotions has been widely studied in psychology and design. Basic features such as hue, brightness, and saturation significantly influence how people perceive emotions~\cite{oberfeld2015effects, Valdez1994EffectsOC}. Kobayashi~\cite{Kobayashi1992ColorIS} established a system linking color palettes to emotional words. Further research refined ways of linking color and emotion, such as Huang's Lab distance method associating palettes with specific emotions~\cite{Huang2018AutomaticIS}, and Zhang's color-emotion circumplex based on Russell’s affective dimensions~\cite{Zhang2024EmotionalLI}. 
Cross-cultural studies have also found that people from diverse backgrounds often react to colors in similar emotional ways~\cite{Ou2004ASO}.
% Some research in computer vision has utilized color features to predict emotions from images. Machajdik and Hanbury~\cite{Machajdik2010AffectiveIC} trained classifiers using hue distributions. Wedolowska et al.~\cite{Wedolowska2023PredictingEF} demonstrated that color palettes derived from visual content effectively reveal emotional contexts.

\begin{figure*}[t]
	\centering
	\includegraphics[width=0.85\linewidth]{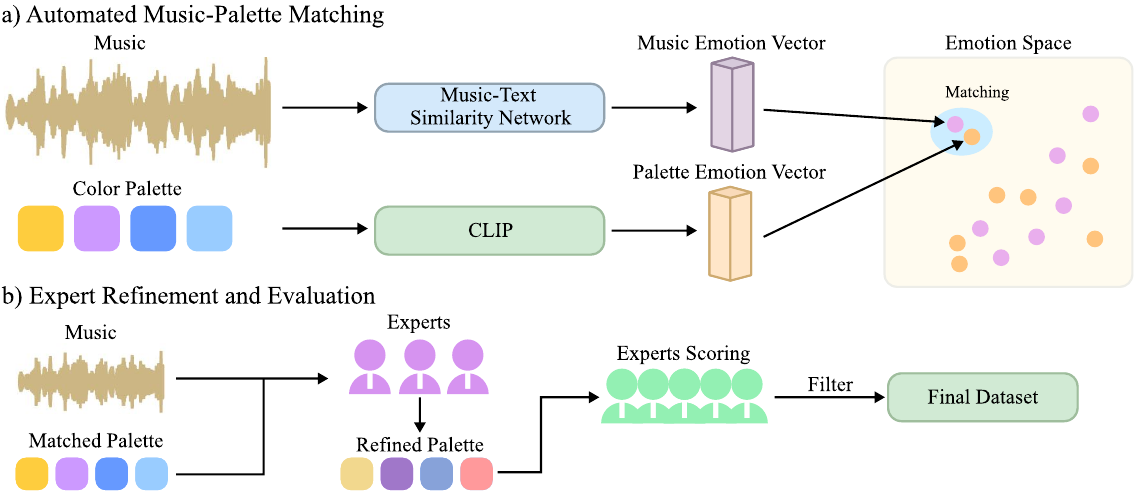}
	\caption{An overview of MuCED. (a) Music and palette pairs are automatically matched based on their emotion vectors in a shared emotion space. (b) Human experts refine and evaluate these pairs to ensure emotion coherence and visual quality.}
	\label{Dataset}
\end{figure*}
% To better reflect emotions by color, researchers are now trying to generate palettes based on emotions.
% Traditional methods like K-means clustering~\cite{ahmed2020generalized,DBLP:journals/tog/Finkelstein15}  and saliency-based extraction~\cite{DBLP:conf/ei-imawm/0002VA15,liu2022image} can extract colors from images, but they do not consider emotional meaning. 
% % Csurka et al.~\cite{Csurka2010LearningMA} tried to connect image features with emotion labels, but their method depends on labeled data.
% Text-driven models like Text2Colors~\cite{DBLP:conf/eccv/BahngYCPWMC18} and Text2Palette~\cite{lei2021text2palette} leverage GANs and Transformers to map emotional keywords to palettes. Moussa et al.~\cite{Moussa2021GenerationAE} employed a VAE to sample from a continuous space and generate color palettes. Inspired by these works, we include a Transformer Color Decoder in our music-driven framework so that we can directly produce vibrant color palettes from audio features. By modeling color sequences in parallel, this method both maintains emotional consistency and boosts the overall variety of the generated palettes.
To better reflect emotions by color, researchers are now trying to generate palettes that carry emotion.
Methods like K-means clustering~\cite{ahmed2020generalized,DBLP:journals/tog/Finkelstein15}  and saliency-based extraction~\cite{DBLP:conf/ei-imawm/0002VA15,liu2022image} can extract colors from images, but they cannot reflect specific emotions. 
% Csurka et al.~\cite{Csurka2010LearningMA} try to connect image features with emotion labels, but their method depends on labeled data.
% These challenges have led to the adoption of deep learning methods for more expressive palette generation.
Recent approaches turn to deep learning for more expressive palette generation.
Text-driven models like Text2Colors~\cite{DBLP:conf/eccv/BahngYCPWMC18} and Text2Palette~\cite{lei2021text2palette} leverage GANs and Transformers to map emotional keywords to palettes. Moussa et al.~\cite{Moussa2021GenerationAE} employed a VAE to sample from a continuous space and generate color palettes.
Inspired by these works, we include a Transformer Color Decoder in our music-driven framework so that we can directly produce vibrant color palettes from audio features. 
% By modeling color sequences in parallel, this method both maintains emotional consistency and boosts the overall variety of the generated palettes.

\subsection{Multi-Modal Emotion Modeling}
With the rise of multimodal learning, music has become a key bridge between visual and auditory modalities due to its emotional expressiveness, driving research in emotion-driven cross-modal generation and alignment. Li et al.~\cite{Lee2020CrossingYI} first use GANs to map music emotion to visual styles, while methods like MeLFusion~\cite{Chowdhury2024MELFuSIONSM} and Diff-BGM~\cite{Li2024DiffBGMAD} use diffusion models for cross-modal generation from visual content to music, highlighting strong emotion interactions between these two modalities.

To enhance the capture of cross-modal emotional associations, many studies utilize the Valence-Arousal space as a cohesive representation framework.
Zhao et al.~\cite{Zhao2020EmotionBasedEM} and Stewart et al.~\cite{Stewart2023EmotionAlignedCL} demonstrate higher alignment accuracy between music and images with end-to-end networks and contrastive learning . Additionally, Nakatsuka et al.~\cite{Nakatsuka2023ContentBasedMR} propose a cross-modal emotion memory system using feature embedding to improve the robustness of emotional retrieval. Both Kundu et al.~\cite{Kundu2024EmotionGuidedIT} and Wang et al.~\cite{Wang2024ContinuousEI} make use of emotion representations, particularly through image-to-music generation that results in a higher level of aesthetic quality and emotion coherence. 
Meanwhile, methods such as Art2Mus~\cite{Rinaldi2024Art2MusBV}, VMB~\cite{Wang2024MultimodalMG}, and EmoMV~\cite{Thao2022EmoMVAM}, use pretrained large models combined with semantic bridging~\cite{huang2025ccsumsp}. This improves expressiveness and controllability in emotion modeling by allowing bidirectional generation between music and visuals at fine-grained emotional levels.

Although emotion matching between music and images has advanced, color, a more basic visual carrier of emotion, has rarely been investigated. Specifically, the emotion link between music and palettes lacks systematic study. To fill this gap, we propose the first music-to-palette cross-modal dataset and design an end-to-end model to directly generate emotionally consistent color palettes.

\section{Dataset}

In this section, we introduce MuCED, the first cross-modal dataset linking music emotions and color palettes. To achieve emotion consistency and cross-cultural applicability, the dataset is constructed through an automated emotion-matching procedure, which is further refined by human experts.
~\autoref{Dataset} shows the overall pipeline for creating our dataset. The process consists of three main steps. First, we collect two modalities of data, music and palettes, from public datasets (Section~\ref{sec:data-collection}). Next, an automated matching algorithm is used to align the two modalities in a shared emotion space (Section~\ref{sec:automated-music-palette-matching}). Finally, expert refinement and evaluation processes are conducted to guarantee the quality of the dataset (Section~\ref{sec:expert-calibration-refinement}).

\subsection{Data Collection}
\label{sec:data-collection}
The music corpora used in the MuCED are drawn from three public music emotion datasets, namely, Emotify~\cite{aljanaki2016studying}, PMEmo~\cite{Zhang2018ThePD}, and DEAM~\cite{alajanki2016benchmarking}. The Emotify dataset has 400 music tracks annotated with the GEMS, which classifies emotions into nine discrete categories. The DEAM dataset contains 1,802 tracks, and the PMEmo dataset contains 794 tracks. Both datasets employ continuous emotion annotations based on the valence-arousal model. 
To ensure consistent audio quality, all audio files are stored at a sampling rate of 22,050 Hz. While the majority of the music excerpts have a duration of 45 seconds, the remaining samples vary in length.

In terms of palette data, we collect 5,992 unique palettes from Color Hunter~\footnote{https://colorhunt.co/}, Color Lovers~\footnote{https://www.colourlovers.com/}, and Kobayashi’s Color Image Scale datasets~\cite{Kobayashi1992ColorIS}, with each consisting of 3 to 5 colors. To enhance visual diversity and support cross-modal alignment, we de-duplicate palettes by sorting the relative order of Lightness (L), Chroma (C), and Hue (H) in the CIELCh color space. 
% This process ensures richness in visual perception and reduces redundancy in the data.

\subsection{Automated Music-Palette Matching}
\label{sec:automated-music-palette-matching}
To construct a cross-modal dataset based on emotion alignment, we first employ an automated method for initial matching. The key idea is to map music and color palettes into a shared emotion space and measure their similarity based on emotion vectors. This approach allows us to efficiently identify music-palette pairs that are most similar in terms of emotion attributes, providing a solid foundation for subsequent manual refinement.

\textbf{Russell's Circumplex Model.} In this work, we employ Russell's circumplex model~\cite{Russell1980ACM} to generate emotion vectors that capture the emotion of music and color palettes. The model defines emotions on two core dimensions, valence and arousal, which categorize emotions into eight basic categories, excited, happy, content, calm, depressed, sad, afraid, and angry. These categories are used as emotion tags to align music and color palettes with a shared emotion space. Russell's circumplex model has been widely applied in both music emotion analysis~\cite{Zhang2018ThePD, alajanki2016benchmarking} and color psychology~\cite{Nijdam2005MappingET} and, as a result, it is well-suited for cross-modal applications. Moreover, as Russell’s model has been extensively used in vast linguistic corpora, it has high applicability to music-text alignment model and text embeddings~\cite{Nandwani2021ARO}.

\textbf{Emotion Vector.} Based on the eight emotion categories defined by Russell's circumplex model, we propose emotion vectors that capture the emotion properties of music and color palettes as an intermediary for aligning music and color palettes in a shared emotion space. According to this, we define the emotion vectors $E_{music}$ and $E_{palette}$ as follows:
\begin{align}
E_{music} &= \{ y_{e}, y_{h}, y_{co}, y_{ca}, y_{d}, y_{s}, y_{af}, y_{an} \}, \\
E_{palette} &= \{ x_{e}, x_{h}, x_{co}, x_{ca}, x_{d}, x_{s}, x_{af}, x_{an} \},
\end{align}
where $y_i$ and $x_i$ represent the similarity scores that reflect the emotional correlation of music or palettes to each emotion category, reflecting their emotional distribution in Russell’s circumplex space.
To map music and palettes into this emotion space and compute these vectors, we use the following specialized models. For music, we employ a pre-trained music-text  model~\cite{Doh2022TowardUT}, which uses contrastive learning to align corresponding music embeddings with text embeddings. For palettes, we use the text-embedding-3-large model~\cite{textEmbedding} to compute their semantic similarity between palettes and emotion descriptors.

\textbf{Emotion Alignment.} Having obtained the emotion vectors for the music and the color palettes, we assess their affective coherence by computing the cosine similarity between the emotion vectors. Specifically, the similarity between the music emotion vector $E_{music}$ and the palette emotion vector $E_{palette}$ is defined as:
\begin{align}
S(E_{music}, E_{palette}) = \frac{E_{music} \cdot E_{palette}}{\|E_{music}\| \|E_{palette}\|},
\end{align}
where the similarity ranges between -1 and 1, with higher values indicating more emotion consistency between music and palette. Based on this measure, we determine music-palette pairs that demonstrate maximum affective consistency, with the selected color schemes faithfully capturing the emotional character in the associate pieces.

\subsection{Expert Refinement and Evaluation}
\label{sec:expert-calibration-refinement}
To ensure cross-modal consistency, emotion alignment, visual compatibility, and cultural robustness, we design a two-stage expert validation process, including color refinement and evaluation. A total of 45 experts, predominantly from the art and design domain, participated in this process. In the first stage, 20 experts focus on initial color adjustments, while in the second stage, 25 experts conduct subsequent evaluations to mitigate subjective biases and cultural variations.

\textbf{Evaluation Criteria.} In the construction and evaluation of color palettes, we establish a set of criteria to guide experts in generating a dataset that faithfully captures the music-palette relationship. 
To ensure a formal and systematic evaluation, we define three key criteria: 1) \textit{Emotion Consistency}, providing a cross-modal assessment mechanism to evaluate whether the emotion attributes of the palette align with those of the music, where emotions are determined based on Russell’s circumplex model of affect; 2) \textit{Relationship with Musical Feature}, correlating music properties, pitch, loudness, and rhythm, with color properties such as brightness, saturation, and contrast, capturing the dynamic nature of music~\cite{Palmer2013MusiccolorAA}; 3) \textit{Color Complexity and Diversity}, maintaining a rich mix of shades, intensities, and balance to create expressive and perceptually aligned palettes that reflect the mood of the music. 
These criteria serve as fundamental principles, ensuring that experts generate a dataset where music-palette relationships remain expressive, diverse, and perceptually aligned throughout the construction process.

\textbf{Expert Refinement.} In the first stage, we first use an automatic matching algorithm to recommend the five most similar palettes for each piece of music.
The palettes are edited independently by three experts. During this adjustment, the experts strictly follow the evaluation criteria, considering the music emotion, its structural characteristics, and the visual coherence of the colors to refine the palettes. An adjustment can involve subtle changes, such as hue shifts, re-calibration of the saturation of the colors, or even complete re-arrangement of the color configuration, to better fit with the music's emotion representation. Each expert independently refines their palettes, resulting in three distinct versions for each piece of music. All refined palettes are kept and taken to the second step of evaluation.

\textbf{Expert Evaluation.} In the second stage, each music-palette pair is rated independently by at least 10 experts. To minimize the potential biases brought on by cultural background and individual preferences, each sample will be evaluated by at least 3 experts from different cultures. 
% Randomized assignment will allow for the even distribution of the workload while keeping one expert from over-scoring a particular piece. 
A five-point Likert scale is used for the scoring process, where 1 indicates a poor match and 5 indicates a strong match. The evaluation takes into account the palette's visual attractiveness and overall coherence in addition to its alignment with the music's emotional qualities. The final dataset does not include pairs with an average score of less than 3.5. If more than one palette is rated highly for the same piece of music, the highest-rated version of the music is incorporated into the final dataset. 
% This procedure reduces the impact of biases between cultural differences and diminishes subjective contributions from individual palettes associated with colors independently reviewed with respective experts, thereby increasing the overall quality and objectivity of the dataset.

After expert validation, we retained 2,634 music-palette pairs with strong emotion alignment, achieving an average similarity score of 0.76. These pairs also exhibited cross-cultural adaptability and visual diversity, reflected in an average expert rating of 4.36.

\section{Method}
% \begin{figure*}[t]
% 	\centering
% 	\includegraphics[width=\linewidth]{figs/MainPipeline.png}
% 	\caption{An overview of Music2Palette.}
% 	\label{Pipeline}
% \end{figure*}

\begin{figure}[t]
	\centering
	\includegraphics[width=0.9\linewidth]{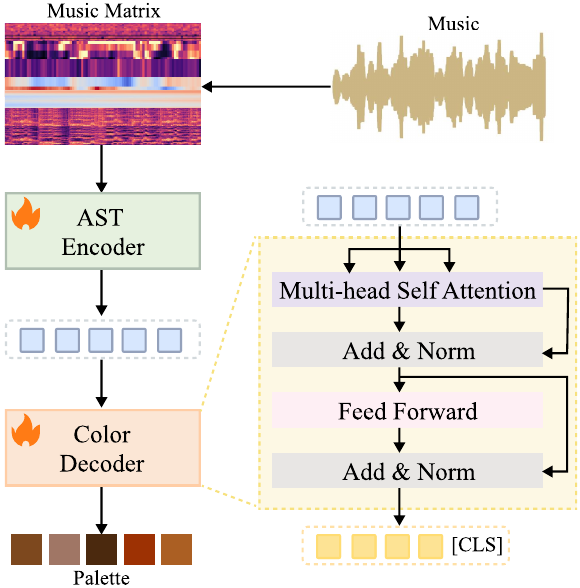}
	\caption{An overview of Music2Palette.}
	\label{Pipeline}
\end{figure}
% In this section, we introduce our proposed method for music-driven palette generation.
In this section, we propose a cross-modal representation learning framework that maps music into a shared emotion space and decodes it into emotion-aligned color palettes.
The framework is shown in~\autoref{Pipeline}. 
First, an improved AST encodes the Mel spectrogram along with rhythm, pitch, and other acoustic features into embeddings. This captures the emotion changes across time and frequency. 
To enhance the diversity of generated palettes, we then introduce Gaussian noise into the music embedding space to encourage diverse color outputs. 
Then, a Transformer-based color decoder generates color palettes of different lengths. It uses multi-head attention to pay attention to the music embeddings during decoding, allowing each color to reflect emotional cues in the music while maintaining coherence within the palette.
To balance the competing goals of visual coherence and emotion expressiveness, we propose a multi-objective loss that jointly optimizes color coherence, palette diversity, and emotion alignment.
This joint training strategy ensures visually appealing and emotionally faithful results.
\subsection{Architecture}
\textbf{Music Feature Processing.} Recent music models~\cite{dong2024musechat, gong2021ast} predominantly rely on the Mel spectrogram to represent music. However, previous studies have shown that both emotional~\cite{lin2024emotional} and acoustic characteristics~\cite{palmer2016music} influence the relationship between music and color, including chroma, spectral contrast, tonal centroid, tempo, energy, and pitch. Inspired by these studies, we incorporate additional music features and align them with the Mel spectrogram along the temporal dimension. The aligned features are then standardized and concatenated to form a unified feature matrix, which is subsequently fed into the Music Encoder for representation learning.

\textbf{Music Encoder.} Our encoder is an AST (Audio-Spectrogram-Transformer)~\cite{gong2021ast}, but adapted to process concatenated features. In the standard AST, it encodes fixed-length mel-spectrogram inputs using a learnable positional embedding. However, since our input consists of concatenated segments, the fixed-length assumption no longer holds. To address this, we modify the positional encoding mechanism of AST by applying sinusoidal encoding exclusively along the temporal dimension \(W\), which is defined as follows:

\begin{equation}
    PE(pos, i) =
    \begin{cases}
        \sin\left(\frac{pos}{10000^{2i/d}}\right), & i \text{ is even}, \\
        \cos\left(\frac{pos}{10000^{2i/d}}\right), & i \text{ is odd},
    \end{cases}
\end{equation}
where \(pos\) represents the position index, \(i\) is the feature dimension index, and \(d\) is the dimension of the encoding vector. Compared to the original AST method, this revised positional encoding strategy eliminates the need for interpolation, making it compatible with variable-length inputs. 
% This design is intended to better capture temporal structures while maintaining efficiency.

\textbf{Embedding Augmentation.} The mapping from music to color palettes is diverse and subjective, as one music can evoke numerous diverse color palettes based on individual perception. A deterministic technique would produce extremely identical color palettes, restricting palette diversity. To address this, we add controlled Gaussian noise to the music embeddings:
\begin{equation}
% z_{\text{aug}} = z_m + \epsilon, \quad \epsilon \sim \mathcal{N}(0, \sigma^2I)
z_{\text{aug}} = z_m + \epsilon, \quad \epsilon \sim \mathcal{N}(0, I),
\end{equation}
where $z_m \in \mathbb{R}^{d_m}$ represents the original music embedding. This approach creates a local variation space around each embedding, allowing the decoder to produce different but emotionally consistent color palettes.

\textbf{Color Decoder.} A transformer decoder is employed for generating the color palette. With multi-head self-attention, the decoder effectively extracts temporal information from music embeddings to provide stylistic consistency. Given that certain musical attributes correlate with human color perception~\cite{bresin2005color}, we represent colors in the CIELCh color space. Furthermore, to accommodate the different complexity and expressiveness of different musical works, we introduce a learnable stop token \texttt{[CLS]}, which dynamically regulates the output length, allowing the palette to adapt to different music contexts without arbitrary size constraints flexibly.

\subsection{Loss Design}
% \subsection{Multi-Objective Color Alignment Loss}
In the music-guided color palette generation task, we seek to enhance the diversity of generated palettes and maintain emotion consistency with the music. To achieve this, we propose a multi-objective loss function composed of \textit{color distance loss, color diversity loss, and emotion consistency loss} to optimize model performance.
% In the music-guided color palette generation task, we seek to enhance the diversity of generated palettes and maintain emotion consistency with the music. To achieve this, we propose a multi-objective loss function, Multi-Objective Color Alignment Loss, consisting of \textit{color distance loss, color diversity loss, and emotion consistency loss}. These objectives collaboratively optimize both perceptual and affective alignment.

\textbf{Color Distance Loss.} To measure the color proximity between the generated and ground-truth palettes, we use the CIEDE2000~\cite{sharma2005ciede2000} in the LCH color space. Since the generated colors may not have explicit correspondences with the ground-truth colors, we employ the Hungarian algorithm~\cite{Munkres1957ALGORITHMSFT} to determine the optimal color arrangement \(\hat{\pi}\) that minimizes the overall color distance:
\begin{equation}
\hat{\pi} = \mathop{\arg\min}_{\pi \in \mathcal{P}} \sum_{i=1}^{N}\Delta E_{00}(c_i,\hat{c}_{\pi(i)}),
\end{equation}
where \( c_i \) represents the colors in the ground-truth palette, \(\hat{c}\) represents the generated palette colors, and \(\mathcal{P}\) represents the set of all possible color permutations. \(\Delta E_{00}\) represents the CIEDE2000, a perceptually weighted formula that evaluates differences in LCH color space. Thus, the color distance loss is defined as:
\begin{equation}
L_{\text{color}} = \frac{1}{N}\sum_{i=1}^{N}\Delta E_{00}(c_i,\hat{c}_{\hat{\pi}(i)}),
\end{equation}

\textbf{Color Diversity Loss.} To encourage diversity in the generated palettes and prevent color repetition, we introduce a diversity loss that penalizes similar hues within the same palette. Specifically, we calculate the Euclidean distances between all colors in the hue channel and add a penalty when colors are too close to each other. The color diversity loss is defined as:
\begin{equation}
L_{\text{diversity}} = -\frac{1}{N(N-1)}\sum_{i\neq j}\|h_i - h_j\|_2,
\end{equation}
where \( h_i, h_j \) are the hue values of distinct colors in the generated palette, and \(N\) is the number of colors per palette.

\textbf{Emotion Consistency Loss.} To ensure the generated palette aligns with the music emotion, we measure emotion consistency by computing the cosine similarity between the emotion vectors of the music \(E_m\) and the predicted palette \(E_p\). Furthermore, to reinforce alignment with the ground truth palette, we compute the similarity between the emotion vectors of the generated palette \(E_p\) and ground-truth palette \(E_g\). The emotion consistency loss is defined as:
% \begin{equation}
% L_{\text{emotion}} = 2 - \text{CosSim}(E_m, E_p) - \text{CosSim}(E_p, E_g)
% \end{equation}
\begin{equation} 
L_{\text{emotion}} = 2 - \frac{E_m \cdot E_p}{|E_m| |E_p|} - \frac{E_p \cdot E_g}{|E_p| |E_g|}
\end{equation}

Combining all three components together, the total loss function can be represented as:
\begin{equation}
L_{\text{total}} = \lambda_{\text{color}} L_{\text{color}} + \lambda_{\text{diversity}} L_{\text{diversity}} + \lambda_{\text{emotion}} L_{\text{emotion}}
\end{equation}

\section{Application}
\subsection{Music-driven Image Recoloring}
In image generation and recoloring, style control is crucial for conveying emotion, especially in music-related scenarios.
Unlike vision or language, music naturally conveys emotion shifts and atmospheric qualities over time. These characteristics make it an intuitive control signal for generating emotionally aligned color palettes in images.
We propose a music-driven palette generation method that bridges auditory and visual modalities. By recoloring images with music-conditioned palettes, our method enables accurate emotion-to-style translation. This approach offers a structured, interpretable control mechanism and enhances multimodal generation systems' emotion alignment and artistic expressiveness.

\begin{figure}[t]
	\centering
	\includegraphics[width=\linewidth]{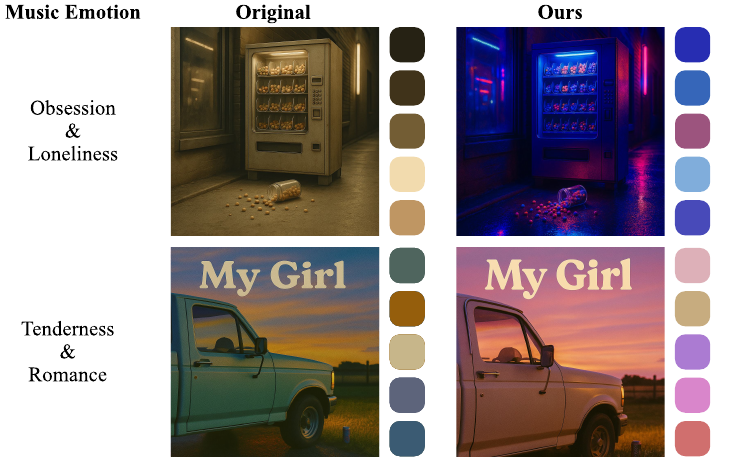}
	\caption{Results of applying music-driven color palettes to images. The palettes are generated based on the input music
and then used to stylize the original images.}
	\label{Application1}
\end{figure}

As shown in ~\autoref{Application1}, our method captures music emotion and guides image recoloring for stronger audio-visual alignment. 
In the scene depicting a candy vending machine, the original music expresses sweetness and a flirtatious nature with uncertainty. To capture this nature, the old, dull brown tones are exchanged for neon tones, showcasing a dreamlike city-night scene of color and light and darkest shades.
% Similarly, in ~\autoref{Application1} (b), an old pickup truck shifts from cold tones to warm pinks, conveying the mood of youthful love expressed in the song. 
Similarly, an old pickup truck shifts from
cold tones to warm pinks, conveying the mood of youthful love
expressed in the song.

\begin{figure*}[t]
	\centering
	\includegraphics[width=\linewidth]{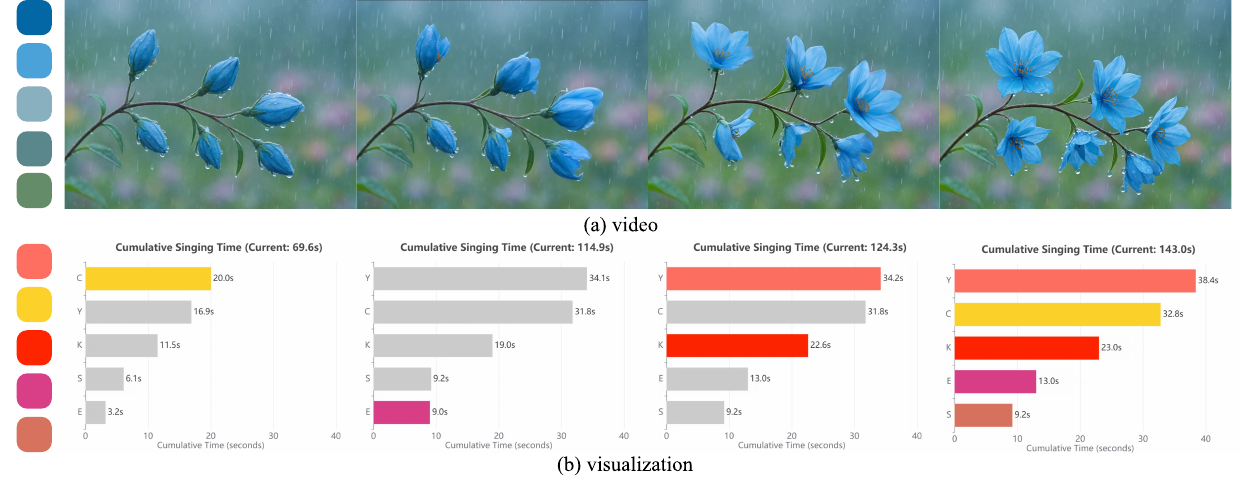}
	\caption{Applications of music-driven color palettes to video and data visualization. (a) In the video, the music conveys a quiet, peaceful feeling. (b) In the visualization, the music reflects a pop song about confidence and bravery.}
	\label{Application2}
\end{figure*}

\subsection{Music-driven Video Generation}
% In video generation, color palettes help connect the emotions in music to what we see on the screen. This makes it easier to control the style of the video and keep it consistent. Instead of using music to directly control the visuals, we first turn music features into matching colors. These colors show the same feelings as the music and guide the overall look of the video. The palette also keeps the video from changing styles too suddenly. 
% % It works well for videos with many scenes or big changes in emotion. 
% Users can even adjust the palette to better show the feelings they want, making the video more expressive.
In video generation, achieving harmony between music and visuals is crucial for crafting an immersive experience. Color plays a key role in conveying visual emotion, setting the tone, and deepening the emotion link with music. Yet, most current approaches rely heavily on text as the main guide, which often lacks the precision to describe detailed visual styles, especially subtle shifts in emotion. Music, by contrast, carries rich, time-varying emotion, making it a strong complement to the text. To improve emotion consistency and style alignment, we generate palettes from music and use them to guide the overall color style of the video. This method makes the visuals more vivid and closely aligned with the music emotion.

As shown in~\autoref{Application2} (a), our generated palette is used on a short scene where blue flowers slowly bloom under rain. The background music conveys a quiet and peaceful mood, like raindrops softly falling on petals. To match this mood, our palette is built with soft blues and greens, expressing stillness and natural beauty.

\subsection{Music-driven Data Visualization}
Music introduces emotion elements into data visualizations, enhancing engagement levels and meaningful interpretation. 
Aligning palettes with the rhythm and mood of music allows visualizations to go beyond basic factual displays and build story-like experiences.
% , data representations can transcend basic factual displays to build story-like experiences.
This is especially valuable in mood-tracking systems, where users upload music and text to receive emotion-aware visual feedback. In interactive storytelling, music shapes the reading rhythm and atmosphere. However, if the color palette does not align with the music emotion, it can break the sense of immersion. 
Music2Palette takes music as input and generates palettes based on its emotion changes, helping visual charts express emotions more clearly.

As shown in~\autoref{Application2} (b), our method is used to visualize lyric part statistics in a girl group song. We represent each member’s singing duration over time with a dynamic bar chart. The generated warm and saturated palette reflects the song’s themes of bravery and confidence. This makes it easier to tell the members apart and gives the whole chart a bold and empowering emotion tone.

% As a result, the audience can better understand the structure of the lyrics and the emotion behind the music.

\section{Experiments}

\subsection{Implementation Details}
In this study, we use the Python library librosa~\cite{McFee2015librosaAA} to extract music features, including 128 Mel-spectrogram features, 12 chroma features, 7 spectral contrast features, 6 tonal centroid features, 384 rhythm features, 1 RMS energy feature, and 1 pitch feature, totaling 539 dimensions.
To encode these music features, we utilize the AST model~\footnote{The pretrained weight used is MIT/ast-finetuned-audioset-10-10-0.4593 for the AST}. The Color Decoder consists of 4 Transformer layers, each containing 1 multi-head self-attention module (with 8 heads) and a feedforward neural network. It creates variable-length sequences, with each color created by a feedforward layer with LayerNorm, followed by a linear projection and a Sigmoid activation function. The dropout rate is set to 0.1.
The model is trained on a dataset of 2,634 music samples paired and their corresponding palettes, with 80\% used for training, 10\% for validation, and 10\% for testing. We use the AdamW optimizer with an initial learning rate of $1\times10^{-4}$, then use a CosineAnnealingLR scheduler to reduce the learning rate gradually.
Training is conducted on an NVIDIA GeForce A6000 GPU with 48GB of memory for 100 epochs, with a batch size of 16. 
% The weight decay coefficient for AdamW is set to 0.1.

\subsection{Compared Baselines}
In this section, we set up several baseline models to assess our proposed method in the music-driven palette generation task. We introduce two types of baselines, text-based methods and image-based methods. Each baseline is denoted as A, B, or C for clarity in subsequent quantitative comparisons.
% We use our approach to provide these baselines and assess them using various metrics.

\textbf{Text-driven Baseline Models.} (A) \textit{LP-MC + Text2Colors}: This baseline uses LP-MC~\cite{doh2023lp} to analyze the music content and generate textual descriptions containing key attributes such as instrument, style, and mood. We extract essential keywords from these descriptions and feed them into Text2Colors~\cite{DBLP:conf/eccv/BahngYCPWMC18} to generate color palettes that reflect core musical features.
(B) \textit{LP-MC + GPT-o1}: Similarly based on LP-MC, this variant inputs the full textual descriptions to GPT-o1~\cite{GPT4V}, which generates palettes aligned with the inferred emotional and stylistic characteristics.

\textbf{Image-driven Baseline Model.} (C) \textit{CDCML + Color Thief}: Following~\cite{Zhao2020EmotionBasedEM}, we compute the emotion similarity between music and images within the Valence-Arousal space and match each music clip to an image with similar affective attributes. Then, Color Thief\footnote{http://lokeshdhakar.com/projects/color-thief/} extracts a five-color palette from the matched image.
% using median-cut quantization.
\begin{table}
  \centering
  \caption{Quantitative comparison of different methods in music-to-palette generation. 
  % Metrics include Diversity (Div.), Multimodality (Multi.), Convex Hull Overlap (CHO), Bhattacharyya Coefficient (BC), Emotion Similarity (ES), and JS Divergence (JS).
  }
  \begin{tabular}{ccccccc}
  \toprule
  \textbf{Method} & \textbf{Div.$\uparrow$} & \textbf{Multi.$\uparrow$} & \textbf{CHO$\downarrow$} & \textbf{BC$\downarrow$} & \textbf{ES$\uparrow$} & \textbf{JS$\downarrow$} \\
  \midrule
  A~\cite{doh2023lp, DBLP:conf/eccv/BahngYCPWMC18} & 19.36 & 6.58 & 0.24 & 0.91 & 0.61 & 0.38 \\
  B~\cite{doh2023lp, GPT4V} & 26.88 & 7.76 & 0.15 & 0.78 & 0.66 & 0.36 \\
  C~\cite{Zhao2020EmotionBasedEM} & \textbf{31.13} & 9.62 & 0.08 & 0.77 & 0.50 & 0.43 \\
  Ours & 28.91 & \textbf{11.12} & \textbf{0.07} & \textbf{0.76} & \textbf{0.74} & \textbf{0.33}\\ \midrule
  GT & 30.39 & - & - & - & 0.76 & 0.32\\
  \bottomrule
  \end{tabular}
  \label{tab:music_palette_diversity}
\end{table}

% \subsection{Evaluation Metics}
\subsection{Diversity Evaluation}
To assess the diversity and creativity of palettes generated by different methods, we utilize four objective metrics:
\textit{Diversity}, \textit{Multimodality}, \textit{Convex Hull Overlap}, and \textit{Bhattacharyya Coefficient}. 
% Together, these metrics assess color variation, generative flexibility, and distributional similarity.

Diversity (Div.) measures the perceptual differences among colors within a single palette. Multimodality (Multi.) assesses the model’s ability to generate multiple distinct palettes for a single music input. 
Convex Hull Overlap (CHO) evaluates the spatial similarity between two palettes by computing the intersection-over-union of their convex hulls in the LCH space.
Bhattacharyya Coefficient (BC) measures distributional similarity between color palettes by comparing their normalized histograms in the LCH space.

The results in Table~\ref{tab:music_palette_diversity} show that our method can create diverse color palettes from music, and these palettes match the feeling or mood of the music well. While the image-driven method slightly outperforms in diversity scores, it often fails to preserve alignment with the music emotion. In contrast, our method achieves a better trade-off between visual diversity and emotion-aware generation.

\subsection{Emotion Similarity Evaluation}
To assess the emotion consistency between the generated color palettes and the music input, we use two objective metrics: \textit{Emotion Similarity} and \textit{Jensen-Shannon Divergence}.
Emotion Similarity (ES) measures the emotion consistency between the music and color palettes by emotion vectors. 
% Higher emotion similarity values indicate the generated palette reflects the emotional qualities of the music input.
JS measures the overall consistency of the overall emotion distributions for the music and color palettes. 
% A lower JS Divergence value demonstrates more consistency in the emotion expressions. 

% \textbf{Emotion Similarity} measures the emotion consistency between the music and color palettes. It is calculated as the cosine similarity between the emotion representations of the music and color palette, both measured in an eight-dimensional affective space and generated from emotion annotations. Higher emotion similarity values indicate the generated palette reflects the emotional qualities of the music input, while lower values reflect the failure of the generated palette to capture the emotion qualities of the input music.
% \textbf{JS Divergence} measures the overall consistency of the overall emotion distributions for the music and color palettes. It is computed as the Jensen-Shannon divergence of the associated emotion representations for each modality. A lower JS Divergence value demonstrates more consistency in the emotional expressions, while a higher divergence indicates the modalities do not measure the same emotion characteristics. 
% This measure allows us to evaluate how much of the affective intent of the music has been preserved in the generated palette.

As shown in Table~\ref{tab:music_palette_diversity}, our method works better at keeping music emotion in the generated color palettes. Text-driven approaches often lose subtle affective cues due to indirect mappings. 
Image-driven methods can capture emotion elements in visuals, but when extracting colors, they often focus on the most eye-catching parts of the image, such as people or objects. These parts might not use colors that match the emotion of the music. As a result, the generated palette may look appealing but fail to convey the right mood.
In contrast, our method learns a direct mapping from music to color, ensuring faithful and coherent emotion representation.
% Image-driven methods also experience significant emotional degradation. During the music-to-image conversion, the pictures may show things like people or objects. These elements express emotions through actions or shapes, not through color. As a result, the color in the image may not match the feeling of the music. Later, when colors are picked from the image, the method usually chooses the most eye-catching parts. These parts may not best represent the mood of the music. Because of this, the final color palette may not fit the emotion of the input music. 
% Our method avoids these problems. It learns to go directly from music to color, and this helps keep the emotion meaning clear and accurate.

\subsection{Emotion Consistency via Image Recoloring}
\begin{table}[t]
  \centering
  \caption{Quantitative comparison of image recoloring results using different palette generation methods.}
  \begin{tabular}{lcccc}
  \toprule
  \textbf{Metric} & \textbf{A~\cite{doh2023lp, DBLP:conf/eccv/BahngYCPWMC18}} & \textbf{B~\cite{doh2023lp, GPT4V}} & \textbf{C~\cite{Zhao2020EmotionBasedEM}} & \textbf{Ours}\\
  \midrule
  CLIP-S     & 0.27 & 0.31 & 0.29 & \textbf{0.35}  \\
  Emo-A    & 29.3 & 33.43 & 19.03  & \textbf{40.24} \\
  Emo-S    & 0.13 & 0.14 & 0.10  & \textbf{0.28} \\
  \bottomrule
  \end{tabular}
  \label{tab:image}
\end{table}
% To further assess the emotional expressiveness of palettes from different methods in image recoloring, we use the same music-image pair as input and apply a unified recoloring framework~\cite{GPT4O} to generate results. 
To achieve stronger emotion alignment between music and visual content, we introduce an application that recolors images using palettes generated from music. Here, we evaluate this approach by comparing the recoloring results of different palette generation methods on the same music-image inputs, using a unified recoloring pipeline~\cite{GPT4O}.
The results are evaluated using three metrics: \textit{CLIP similarity}, \textit{Emotion Accuracy}, and \textit{Emotion Incremental Score}.

CLIP similarity (CLIP-S) measures how well the image aligns with the semantic meaning of the music description.
Emotion Accuracy (Emo-A)~\cite{Yang2024EmoGenEI} uses a pre-trained emotion classifier~\cite{Yang2023EmoSetAL} to check whether the recolored image express the intended target emotion.
Emotion Incremental Score (Emo-S)~\cite{Yang2024EmoEditEE} evaluates how much the recoloring process strengthens the target emotion compared to the original image.
As shown in Table~\ref{tab:image}, our method outperforms all baselines across the metrics, highlighting its superiority in both emotion expressiveness and cross-modal semantic alignment.

\subsection{Subjective User Study}
\begin{table}[t]
  \centering
  \caption{Comparison of user study scores on emotion matching and enhancement between music and palettes.}
  \begin{tabular}{ccccc}
  \toprule
  \textbf{Method} & \textbf{Task1} & \textbf{Task2} & \textbf{Task3} & \textbf{Task4}\\
  \midrule
  A~\cite{doh2023lp, DBLP:conf/eccv/BahngYCPWMC18} & 9.52 & 3.14 & 3.05 & 2.95\\
  B~\cite{doh2023lp, GPT4V} & 14.29 & 3.17 & 3.52 & 3.67 \\
  C~\cite{Zhao2020EmotionBasedEM} & 7.14 & 3.12 & 2.82 & 3.52  \\ \midrule
  \textbf{Ours} & \textbf{69.05} & \textbf{4.15} & \textbf{4.24}  & \textbf{4.38}\\
  \bottomrule
  \end{tabular}
  \label{tab:user_study}
\end{table}
% \vspace{-1em}

To assess the effectiveness of Music2Palette from a perceptual and cognitive perspective, we conducted a user study comparing our method with baselines. In \textbf{Task 1}, users listened to a music clip and then chose the color palette they felt best matched the music emotion. In \textbf{Task 2}, users were given a music and palette pair and rated how well they matched on a scale from 1 to 5. In \textbf{Task 3}, participants first viewed a palette, then listened to the associated music, and rated how much the color helped them understand the music emotion. In \textbf{Task 4}, the order was switched. Users listened to the music first, and then looked at the colors, rating whether the palette made their emotional feeling stronger.
The results in Table~\ref{tab:user_study} show that our method performed better than baselines across evaluation stages. Moreover, in Tasks 3 and 4, showing colors before or after music improved participants’ emotion comprehension.
% We first aim to verify whether the emotion representations predicted by our model accurately reflect human perception in each modality. In \textbf{Tasks 1} and \textbf{Task 2}, participants separately listened to music pieces and looked at color palettes, then categorized them into eight emotion categories in the Russell emotion space. By comparing the user-matched labels to the emotion vectors, we found high recognition accuracies: 88.1\% for music and 83.0\% for palettes. This shows that the emotion vectors effectively capture emotion meaning in both auditory and vision.

% To better assess the emotion connections between modalities, we designed four additional tasks. In \textbf{Task 3}, users listened to a music clip then selected the most emotionally matching palette from options created by both baselines and our method. \textbf{Task 4} asked users to score the emotion consistency of a given music-palette pair using a 5-point Likert scale. In \textbf{Task 5}, participants first viewed a palette, then listened to the associated music, and rated how much the color helped them understand the music’s emotion. \textbf{Task 6} reversed the order: participants listened to music first and then viewed the palette, and evaluated whether the visual colors reinforced their emotional impression.

% This suggests that Music2Palette effectively links sound and color, making it useful for applications that combine music and visuals.

\section{Conclusion}
% In this work, we propose a generative model that produces emotionally aligned color palettes from music.
% To support this, we construct MuCED, a large-scale cross-modal dataset of music-palette pairs aligned and refined based on the Russell emotion model and expert annotations.
% Based on this dataset, our model uses a music encoder and color decoder to generate palettes that reflect the music emotion. Experiments indicate that our approach can produce a diverse and expressive color palette that reflects the mood of the music. Music2Palette can also be applied to use in real-world applications, such as for recoloring images, video generating, and data visualization.
% In the future, our method will be expanded to accommodate user customization and real-time interactions.
% We also intend to continue testing the emotional impact of our created color palettes in various multimedia scenarios and look into a variety of cross-cultural and different music sources.
In this work, we propose a novel framework for generating emotion-aligned palettes from music.
To support this, we construct MuCED, a large-scale cross-modal dataset of music-palette pairs aligned using Russell's emotion model and expert annotations.
Based on this dataset, we develop a generative model with a music encoder and a color decoder to produce the palettes. Experiments indicate that our approach can produce a diverse and expressive palette that reflects the music emotion. Music2Palette is applicable to real-world tasks, such as image recoloring, video styling, and dynamic data visualization.
In the future, our method will be expanded to accommodate user customization and real-time interactions.
We also intend to continue testing the emotional impact of our created color palettes in various multimedia scenarios and look into a variety of cross-cultural and different music sources.

\begin{acks}
The authors wish to acknowledge the support from NSSFC under Grant 22ZD05, NNSFC under Grant 62472178, and the Natural Science Foundation of Shanghai Municipality, China under Grant 24ZR1418300.
\end{acks}

\bibliographystyle{ACM-Reference-Format}
\balance
\bibliography{sample-base}

% \appendix









\end{document}